\documentclass{PoS}
\usepackage{epsfig}
\usepackage{amssymb,amsmath,bm}
\usepackage{ctable}
\usepackage{dcolumn}

\def\be{\begin{equation}}
\def\ee{\end{equation}}
\def\bea{\begin{eqnarray}}
\def\eea{\end{eqnarray}}

\def\as{{\alpha_s}}

\def\R{\mathrm{Re}}
\def\la{\langle}
\def\ra{\rangle}

\def\mR{\mu_{\mathrm{R}}}
\def\mF{\mu_{\mathrm{F}}}

\def\ep{\epsilon}
\def\cm{\mathcal{M}}
\def\cf{\mathcal{F}}

\title{Four-dimensional formulation of the sector improved residue subtraction}

\ShortTitle{Four-dimensional formulation of the sector improved
  residue subtraction}

\author{\speaker{David Heymes}\\
        RWTH Aachen University\\
        E-mail: \email{dheymes@physik.rwth-aachen.de}}

\abstract{We report on recent progress we made towards the four-dimensional 
  formulation of the sector improved residue subtraction. We explain
  how the subtraction scheme STRIPPER, proposed in
  \cite{Czakon:2010td}, is
  generalized to higher multiplicities and therefore furnishes a
  general framework for the calculation of next-to-next-to-leading
  order cross sections in perturbative QCD.}

\FullConference{ Loops and Legs in Quantum Field Theory - LL 2014,\\
		 27 April - 2 May 2014 \\
		 Weimar, Germany }

\begin{document}


\section{Introduction}
Upcoming LHC data is expected to reduce the experimental error on
several observable cross sections. In order to test the boundaries of
the Standard Model, more precise theoretical
predictions  have to be provided. General tools to calculate not only
next-to-leading order (NLO) but also next-to-next-to-leading order
(NNLO) cross sections in perturbative QCD are highly demanded.\\
The hadronic cross section factorizes into parton
distribution functions and the renormalized partonic cross section
\be
\sigma_{\mathrm{H}_1\mathrm{H}_2}(P_1,P_2)=\int\limits_0^1\mathrm{d}x_1\mathrm{d}x_2\,
 f_a(x_1,\mF)\,f_b(x_2,\mF)\, {\hat \sigma}_{ab}(x_1P_1,x_2P_2; \as(\mR),\mR,\mF)\;,  
\ee
where the summation over initial state partons $a,b$ (gluons and massless quarks)
is understood.
For high energy processes the partonic cross section can be calculated 
in perturbative QCD. Its expansion in $\as$ up to
NNLO is 
\be
{\hat \sigma}_{ab}={\hat
  \sigma}^{(0)}_{ab}
+{\hat \sigma}^{(1)}_{ab}
+{\hat \sigma}^{(2)}_{ab}+\cdots,
\ee
where 
\begin{align}
{\hat \sigma}^{(0)}_{ab}&=\int_{n}\mathrm{d}{\hat \sigma}^{\mathrm{B}}_{ab}\;
,\\
{\hat \sigma}^{(1)}_{ab}&=\int_{n+1}\mathrm{d}{\hat
  \sigma}^{\mathrm{R}}_{ab}+\int_{n}\mathrm{d}{\hat
  \sigma}^{\mathrm{V}}_{ab}+\int_{n}\mathrm{d}{\hat
  \sigma}^{\mathrm{F}}_{ab}\; ,\\
\label{eq:NNLOpartonic}
{\hat \sigma}^{(2)}_{ab}&=\int_{n+2}\mathrm{d}{\hat
  \sigma}^{\mathrm{RR}}_{ab}+\int_{n+1}\mathrm{d}{\hat
  \sigma}^{\mathrm{RV}}_{ab}+\int_{n}\mathrm{d}{\hat
  \sigma}^{\mathrm{VV}}_{ab}+\int_{n+1}\mathrm{d}{\hat
  \sigma}^{\mathrm{F1}}_{ab}+\int_{n}\mathrm{d}{\hat
  \sigma}^{\mathrm{F2}}_{ab}\, .
\end{align}
The multiplicity of the final state ranges between $n$ and $n+2$
particles. Starting at NLO, ${\hat \sigma}^{(1)}_{ab}$, the calculation suffers from soft and
collinear singularities that emerge as poles in the dimensional
regularization parameter $\ep$, where the dimension is set to 
$d=4-2\ep$. Poles appear after integrating out the
loop momentum of the virtual cross section $\mathrm{d}{\hat  \sigma}^{\mathrm{V}}_{ab}$ or after phase
space integration of the additional parton in the final state of the
real-radiation cross section $\mathrm{d}{\hat
  \sigma}^{\mathrm{R}}_{ab}$. After summing the two contributions
and the collinear factorization counterterm $\mathrm{d}{\hat
  \sigma}^{\mathrm{F}}_{ab}$, poles cancel.
At NNLO, the cancellation between real and
virtual poles continues, but is more involved. It takes place between
the double-real cross section $\mathrm{d}{\hat  \sigma}^{\mathrm{RR}}_{ab}$, where two additional 
partons in the final state can become unresolved, the real-virtual
cross section $\mathrm{d}{\hat  \sigma}^{\mathrm{RV}}_{ab}$, 
that includes the one-loop matrix element and one additional parton 
in the final state, the double-virtual cross section $\mathrm{d}{\hat
  \sigma}^{\mathrm{VV}}_{ab}$, that contains 
two loop matrix elements, and the collinear factorization
counterterms $\mathrm{d}{\hat\sigma}^{\mathrm{F1}}_{ab}$ and 
$\mathrm{d}{\hat  \sigma}^{\mathrm{F2}}_{ab}$ of different multipli\-cities.

Subtraction procedures have been introduced to control the divergent
structure of higher order calculations. The Catani-Seymour
subtraction scheme \cite{Catani:1996vz} and the FKS
(Frixione-Kunszt-Signer) scheme \cite{Frixione:1995ms} are commonly
used at NLO.
At NNLO, Antenna subtraction \cite{GehrmannDeRidder:2005cm} has
 been applied to $e^+e^-\to 3\,\mathrm{jets}$
 \cite{GehrmannDeRidder:2007jk}  and $gg\rightarrow
 2\,\mathrm{jets}$ \cite{Currie:2013dwa}. The $q_{\perp}$-subtraction
 \cite{Catani:2007vq} has been used to calculate Higgs production
 \cite{Catani:2007vq} and $Z$ pair production \cite{Cascioli:2014yka}.
In these subtraction schemes poles cancel analytically as can be seen
by considering the first contributions of equation
\eqref{eq:NNLOpartonic}. The double-real cross section is made
integrable in four dimensions by imposing single
and double unresolved subtraction terms. The real-virtual contribution
is made integrable by suitable single subtraction terms. Its explicit
poles from the virtual amplitude are cancelled analytically by an
integrated subtraction term from double-real radiation. Hence, the $n+1$ particle phase
space integration is completely finite in four dimensions. The
double-virtual contribution has just explicit poles, they are cancelled
by all remaining integrated subtraction terms. Again, this part is
finite in four dimensions. \footnote{For simplicity, the collinear counterterms
in \eqref{eq:NNLOpartonic} were not mentioned in the discussion.}

The subtraction scheme STRIPPER (SecToR ImProved PhasE space for Real
radiation) was originally introduced in \cite{Czakon:2010td} and first
applied to top-quark pair production at NNLO
 \cite{Czakon:2011ve}. Afterwards, it has proven its applicability to other
 processes of low multiplicity: Higgs $+$ jet
 \cite{Boughezal:2013uia},
 charmless bottom quark decay \cite{Brucherseifer:2013cu},
 top quark decay \cite{Brucherseifer:2013iv}, single top quark
 production \cite{Brucherseifer:2014ama},
 muon decay \cite{Caola:2014daa} and Z decay \cite{Boughezal:2011jf}.

Currently STRIPPER is formulated in the conventional dimensional
regularization scheme (CDR), i.e. momenta and spin degrees of freedom
of resolved and unresolved particles are treated in $d=4-2\ep$
dimensions. Unresolved particles are either virtual partons in a
loop or real radiated partons that can become soft or
collinear. In contrast to analytic subtraction schemes
 STRIPPER is a numerical approach to calculate cross
sections. Therefore, a parametrization of momenta has to be provided for all
particles right from the beginning. Using CDR, also unphysical 
dimensions are parameterized, e.g. for
$t\bar{t}$-production already five dimensions had to be parameterized
explicitly \cite{Czakon:2011ve}. Adding an additional particle would
increase the dimensionality by one. Furthermore, all matrix elements that are used for subtraction
terms have to be provided to higher powers in $\ep$. Unfortunately, available
software for leading order matrix elements provide just the four dimensional part. 
The applicability of STRIPPER in CDR for higher multiplicities is
therefore problematic. Hence, it is  desireable to reformulate it in
the 't Hooft-Veltman regularization scheme (HV), where momenta and spin
degrees of freedom of resolved particles are in four dimensions.

The explicit distinction between CDR and HV does not appear in 
analytic  subtraction schemes mentioned above, since all phase
space integrations are four dimensional and the difference between CDR
and HV starts at order $\ep$.

In this proceedings, we explain the general STRIPPER algorithm used to
obtain a Laurent series in $\ep$ for ${\hat
  \sigma}^{\mathrm{RR}}_{ab}$, 
where all coefficients can be integrated
numerically. Subsequently, we explain corrections that have to be
applied in order to reformulate the scheme in HV. 
A general and detailed discussion of the
scheme and all necessary formulas to set up the subtraction in four
dimensions can be found in \cite{Czakon:2014oma}.
%
\section{STRIPPER - Outline of the general procedure}
The double-real radiation contribution to the full cross section has the
explicit form
\be
\label{eq:RR}
{\hat
  \sigma}^{\mathrm{RR}}_{ab}=\int_{n+2}\mathrm{d}{\hat
  \sigma}^{\mathrm{RR}}_{ab}=\frac{1}{2\hat{s}}\int
\mathrm{d}\Phi_{n+2}\langle \cm^{(0)}_{n+2}|\cm^{(0)}_{n+2}\rangle F_{n+2}\;,
\ee
where $\hat{s}$ is the partonic center of mass energy,
$\mathrm{d}\Phi_{n+2}$ the $n+2$ particle phase space and $F_{n+2}$
the measurement function which describes the observable under
consideration. Matrix elements are described as vectors in color and
spin space expanded in $\as$
\be
|\cm_{n}\rangle
=|\cm^{(0)}_{n}\rangle+|\cm^{(1)}_{n}\rangle+|\cm^{(2)}_{n}\rangle+\cdots
\;.
\ee
The squared matrix element appearing in \eqref{eq:RR} is summed
over final state color and spin and averaged over initial state color
and spin. 
Momenta and spin degrees of freedom of particles are treated in
$d=4-2\ep$ dimensions. The integral over the phase space will generate
poles in $\ep$ as one or two partons become unresolved. The following
algorithm provides a systematic approach to calculate \eqref{eq:RR} as
a Laurent series in $\ep$ where all integrals can be integrated
numerically.\\

The first step is to introduce selector functions in order to split
the phase space into double-collinear and triple-collinear sectors
\be
\label{eq:Selector}
1=\sum S_{\mathrm{TC}}+\sum S_{\mathrm{DC}}\;.
\ee
In a triple-collinear sector, $S_{\mathrm{TC}}$, singularities are
generated as three specific partons become collinear to each other
and/or two of them soft. In a
double-collinear sector, $S_{\mathrm{DC}}$, singularities are generated
as two specific pairs of partons become
collinear and/or two of them soft. All remaining soft or collinear limits
are regulated by the selector function, which in general depends on
energies of final state partons and angles between all partons.\\
Selector functions are also used to generate subtraction terms at NLO
and for the real-virtual cross section at NNLO. In these cases, it coincides with the
FKS subtraction method.\\

The second step is to parameterize the collinear particles in each
sector separately using energies and angles. A triple collinear
sector, 
where the three particles are in the final state, is taken as an
example:
The reference momentum that points towards the
triple-collinear direction is denoted by $r^{\mu}$. The momenta of the unresolved
partons are denoted by $u_1^{\mu}$ and $u_2^{\mu}$. Each momentum is
parameterized by its energy and a $(d-1)$-dimensional unit vector
\be
r^\mu \equiv r^0 \begin{pmatrix} 1 \\ \bm{\hat{q}_1} \end{pmatrix}
\; , \quad 
u_1^\mu \equiv u_1^0 \begin{pmatrix} 1 \\ \bm{\hat{u}_1} \end{pmatrix}
\; , \quad
u_2^\mu \equiv u_2^0 \begin{pmatrix} 1 \\ \bm{\hat{u}_2} \end{pmatrix}
\;.
\ee
The unresolved particles' energies are rescaled by their maximal value
$E$, which depends on the considered process, 
$u_i^0={\hat \xi}_iE$, for $i=1,2$. The soft limit is approached as
${\hat \xi_1}\to
0$
and/or ${\hat \xi}_2\to 0$.
The unit vectors are defined by 
\begin{align}
\bm{\hat{r}}\, &\equiv \bm{\hat{n}}^{(d-1)}(\alpha_1, \alpha_2, \dots)
\; ,
\nonumber \\
\bm{\hat{u}_1} &\equiv \bm{R}^{(d-1)}_1(\alpha_1, \alpha_2, \dots)
\bm{\hat{n}}^{(d-1)}(\theta_1, \phi_1, \rho_1, \rho_2, \dots) \; ,
\nonumber \\
\bm{\hat{u}_2} &\equiv \bm{R}^{(d-1)}_1(\alpha_1, \alpha_2, \dots)
\bm{R}^{(d-1)}_2(\phi_1, \rho_1, \rho_2, \dots)
\bm{\hat{n}}^{(d-1)}(\theta_2, \phi_2, \sigma_1, \sigma_2, \dots)
\; ,
\end{align}
where 
\be
\bm{\hat{n}}^{(d-1)}(\alpha,\beta,\dots)=\begin{pmatrix} \vdots \\\sin\alpha
\sin\beta \dots \\ \sin\alpha \cos\beta\\ \cos\alpha \end{pmatrix} \; .
\ee
$\bm{R}^{(d-1)}_1$ and $\bm{R}^{(d-1)}_2$ are $(d-1)$-dimensional
rotation matrices that are chosen in such a way that the
scalar products between the three given momenta take the following
form
\be
\bm{\hat{u}_1} \cdot \bm{\hat{r}} = \cos\theta_1 = 1-2{\hat \eta}_1 \; , \quad
\bm{\hat{u}_2} \cdot \bm{\hat{r}} = \cos\theta_2 = 1-2{\hat \eta}_2 \; , \quad
\bm{\hat{u}_1} \cdot \bm{\hat{u}_2} = \cos\theta_1
\cos\theta_2 + \cos\phi_2 \sin\theta_1\sin\theta_2
\; .
\ee
\begin{floatingfigure}[t]
 \includegraphics[scale=.75]{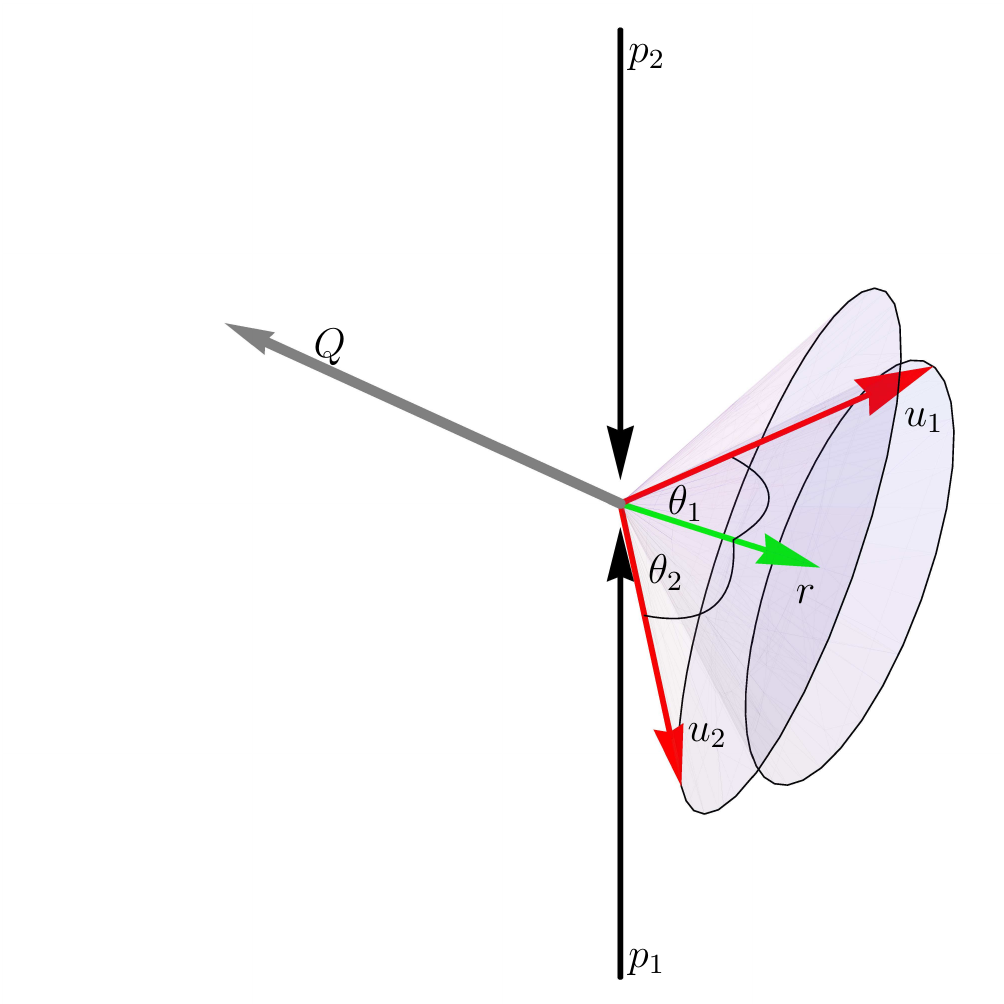}
\caption{\label{fig:Parametrization} Parametrization of a
  triple-collinear sector, where the three partons are in the final
  state. The reference momentum is $r$, the unresolved momenta are
  denoted by $u_1$ and $u_2$. The sum of  momenta of remaining
  resolved particles is $Q$.}
\end{floatingfigure}
This parametrization is depicted in figure \ref{fig:Parametrization}.
The limit of ${\hat \eta}_i$ at zero indicates the collinear limit
of one of the unresolved partons and the reference parton. On the
other hand, $\bm{\hat{u}_1}$ is collinear to $\bm{\hat{u}_2}$ when
${\hat \eta}_1 = {\hat \eta}_2$ and $\phi_2=0$. After transforming
$\phi_2$ non-linearly, it becomes a function of ${\hat \eta}_1$,
${\hat \eta}_2$ and a new integration variable $\zeta$
\be
\phi_2 \rightarrow \phi_2({\hat \eta}_1,{\hat \eta}_2,\zeta). 
\ee
The transformation ensures that $\phi_2$ always vanishes as ${\hat \eta}_1={\hat \eta}_2$. Using this
parametrization just two variables ${\hat \eta}_1$ and ${\hat \eta}_2$ indicate all
possible collinear limits. Furthermore, if a formulation in
HV is found, meaning that all other momenta and the reference momentum are four-dimensional, at most six dimensions are needed to parameterize all
possible scalar products that appear.\\

At this point, four physical variables $\{{\hat \eta}_1,{\hat \eta}_2,{\hat
  \xi}_1,{\hat \xi}_2\}$ parameterize all
possible soft and collinear limits in a given sector. Additional
sector decompositions in those variables factorize all possible
overlapping singularities that appear at NNLO. For example, in the
double-soft limit the phase space integration has the following
schematic  form
\be
S=\int_{0}^{1}{\mathrm{d}\hat{\xi}_1\mathrm{d}\hat{\xi}_2}\,\frac{\hat{\xi}_1^{-2\epsilon}\hat{\xi}_2^{-2\epsilon}}{\left(\hat{\xi}_1+\hat{\xi}_2\right)^2}.
\ee
The integral is singular in the region $\{\hat{\xi}_1,\hat{\xi}_2\}\rightarrow\{0,0\}$.  
Imposing an order on the limits of the soft variables, i.e. splitting
the phase space according to 
\be
\label{eq:Decomp}
1=\theta\left(\hat{\xi}_1-\hat{\xi}_2\right)+\theta\left(\hat{\xi}_2-\hat{\xi}_1\right),
\ee
factorizes the singular limit into one variable 
\be
S=\int_{0}^{1}{\mathrm{d}\xi_1\mathrm{d}\xi_2}\,\frac{\xi_1^{-1-4\epsilon}\xi_2^{-2\epsilon}}{\left(1+\xi_2\right)^2}+\left(1\leftrightarrow2\right),
\ee
where the sector  variables $\{\xi_1,\xi_2\}$ are functions of the
physical variables $\{\hat{\xi}_1,\hat{\xi}_2\}$. \\
The decomposition of
the phase space due to soft overlapping singularities is sufficient to
factorize all possible limits in a double-collinear sector. In a
triple-collinear sector the phase space is split into five additional
sectors to factorize collinear and soft-collinear overlapping
singularities. \\

Finally, the full double-real radiation cross section can be written as a
sum over different sectors that are obtained after introducing
selector functions \eqref{eq:Selector} and applying  sector
decompositions \eqref{eq:Decomp}
\be
{\hat
  \sigma}^{\mathrm{RR}}_{ab}=\sum_{S}{\hat
  \sigma}^{\mathrm{RR},S}_{ab}\; .
\ee
Each contribution has the following form
\be
{\hat
  \sigma}^{\mathrm{RR},S}_{ab}=\int\limits_0^1
\mathrm{d}\xi_1\mathrm{d}\xi_2\mathrm{d}\eta_1\mathrm{d}\eta_2 \frac{F_S\left(\xi_1,
  \xi_2, \eta_1,
  \eta_2\right)}{\xi_1^{1-b_1\epsilon}\xi_2^{1-b_2\epsilon}\eta_1^{1-b_3\epsilon}\eta_2^{1-b_4\epsilon}},
\ee
where all possible singularities are factorized in sector
variables. The function $F_S\left(\xi_1,  \xi_2, \eta_1,
\eta_2\right)$ is finite as the variables go to zero. This form is
appropriate to generate subtraction terms and integrated subtraction
terms, i.e. pole terms, using the definition of the plus distribution
iteratively for each sector variable
\be
\label{eq:Plus}
\int_0^1\mathrm{d}x\,\frac{f\left(x\right)}{x^{1-b\epsilon}}=\frac{f\left(0\right)}{b\epsilon}+\int_0^1\mathrm{d}x\,\frac{f\left(x\right)-f\left(0\right)}{x^{1-b\epsilon}} \;.
\ee
The result is a Laurent series in $\ep$, where all coefficients can be
evaluated numerically.
The described procedure is process independent, since it is possible
to use the known universal IR-limits of QCD amplitudes for the subtraction and
integrated subtraction terms in \eqref{eq:Plus}. The limit of a
vanishing sector variable corresponds to a well defined single or double
unresolved limit of the matrix element in \eqref{eq:RR}.


\section{Reformulation of STRIPPER in four dimensions}

Treating all particles in $d=4-2\ep$ dimensions  ensures that virtual and real poles
arising in the different contributions of \eqref{eq:NNLOpartonic} cancel
after all parts are added. \\
However, the goal is to only treat unresolved particles in $d=4-2\ep$
dimensions and keep all remaining resolved particles in $d=4$
dimensions. Therefore, it is necessary to identify poles of 
different contributions in \eqref{eq:NNLOpartonic} that arise from one unresolved
particle or from two unresolved particles. The sum of pole terms
that arise from one unresolved particle that is proportional to
$\langle \cm^{(0)}_{n+1}|\cm^{(0)}_{n+1}\rangle $ will be called the
single-unresolved contribution.
The sum of all pole terms that arise from two unresolved particles
that is propotional to $\langle \cm^{(0)}_{n}|\cm^{(0)}_{n}\rangle$ 
will be called the double-unresolved contribution. There is one more
type of divergent contribution, those proportional to the finite part of the
$n$-particle one-loop amplitude $\langle \cm^{(0)}_{n}|\cf^{(1)}_{n}\rangle$.
They arise in ${\hat\sigma}^{\mathrm{RV}}_{ab}$,
${\hat\sigma}^{\mathrm{VV}}_{ab}$ and
${\hat\sigma}^{\mathrm{F1}}_{ab}$. It was pointed out that this poles
cancel independent of the single- and double-unresolved contributions
\cite{Weinzierl:2011uz}.  \\
A formulation in the 't Hooft-Veltman regularization scheme means that
in the single-unresolved contribution, $n+1$ particles are in four
dimensions and in the double-unresolved contribution $n$ particles are
in four dimensions. 
This is consistent if it is ensured that these contributions are
finite separately, which is the case only after  corrections of the  
integrated subtraction terms in ${\hat\sigma}^{\mathrm{RR}}_{ab}$. How
these corrections can be identified will be shortly outlined in
section \ref{sec:Modification}. \\
The general procedure to get the transition from CDR to HV is as
follows: First, the identified corrections for the single-unresolved
contribution are added. Then, it is free of poles, i.e. finite in four 
dimensions. Replacing the measurement function $F_{n+1}$ by 
\be
\label{eq:JetFunctionSU}
F_{n+1}\cdot \prod_{i=1}^{n}{\delta^{\left(-2\epsilon\right)}\left( q_i  \right)},
\ee   
restricts $n$ resolved particles of momentum $q_i$ to four
dimensions. Energy-momentum
conservation guarantees that the remaining resolved momentum 
is also four-dimensional. The error induced by \eqref{eq:JetFunctionSU} is of order $\ep$. In addition,
spin-correlated splitting functions are replaced by their azimuthal
avaraged counterparts, if they accompany a collinear pole. 
Finally, momenta and spin degrees of freedom of $n+1$ particles
are four-dimensional and only the four-dimensional part of $\langle \cm^{(0)}_{n+1}|\cm^{(0)}_{n+1}\rangle $ has to be provided.\\
The corrections to the single-unresolved contribution are subtracted
from the double-unresolved contribution and render it 
finite as well. Then, the same arguments as for the single-unresolved
contribution can be applied. The difference is that $n$ particles
are resolved. Therefore, the measurement function
$F_{n}$ is replaced by
\be
\label{eq:JetFunctionDU}
F_{n}\cdot \prod_{i=1}^{n-1}{\delta^{\left(-2\epsilon\right)}\left( q_i  \right)}.
\ee   
After replacing spin-correlated splitting functions by the
azimuthal-averaged ones, only the four-dimensional part of 
$\langle \cm^{(0)}_{n}|\cm^{(0)}_{n}\rangle $ is needed.     
\subsection{Corrections to ${\hat\sigma}^{\mathrm{RR}}_{ab}$ in HV - An example}
\label{sec:Modification}
Here we outline how corrections to the integrated subtraction terms in 
${\hat\sigma}^{\mathrm{RR}}_{ab}$ can be identified, such that the
single-unresolved contribution is finite.
First, it is necessary to identify all three parts of the single-unresolved
contribution in \eqref{eq:NNLOpartonic}.
The real-virtual cross
section is split in a pure pole contribution and a finite remainder
\begin{align}
\label{eq:RVmatrix}
\nonumber
\int_{n+1}\mathrm{d}{\hat  \sigma}^{\mathrm{RV}}_{ab} &= \frac{1}{2\hat{s}}\int
\mathrm{d}\Phi_{n+1}
 2\R\,\la \cm_{n+1}^{(0)} |\cm_{n+1}^{(1)}\ra F_{n+1}\\
 &= \frac{1}{2\hat{s}}\int
\mathrm{d}\Phi_{n+1} \left[2\R\,\la \cm_{n+1}^{(0)}
 |\bm{Z}^{(1)}|\cm_{n+1}^{(0)}\ra +
 2\R\,\la \cm_{n+1}^{(0)} |\cf_{n+1}^{(1)}\ra
 \right]F_{n+1}\;,
\end{align}
where $\bm{Z}^{(1)}$ is an operator in color space and contains 
poles in $\ep$ only. The first term on the r.h.s. of  \eqref{eq:RVmatrix}
is part of the single-unresolved contribution. 
The collinear factorization counterterm adds
\be
\int_{n+1} \mathrm{d}\sigma^{\mathrm{F}_2}_{ab}(p_1,p_2)=\frac{\as}{2\pi}\frac{1}{\ep}\int\limits_0^1\mathrm{d}z
\left[P^{(0)}_{ca}(z) \,\int_{n+1}\mathrm{d}\sigma^{\mathrm{R}}_{cb}(zp_1,p_2) +
  P^{(0)}_{db}(z)\, \int_{n+1}\mathrm{d}\sigma^{\mathrm{R}}_{ad}(p_1,zp_2) \right]\;
\ee
to the single-unresolved contribution, where $P^{(0)}_{ab}(z)$ are the
leading order splitting kernels. In the double-real cross section, 
integrated subtraction terms generated by STRIPPER due to one
unresolved parton contribute to the single-unresolved contribution.\\
If there existed a NLO measurement function in
the $n+1$-particle phase space that prevented all particles from
becoming soft or collinear,  the pole cancellation would be the known
NLO pole cancellation between integrated subtraction terms and virtual
poles. However, at NNLO one add\-itional particle can become
unresolved. Hence, subtraction terms for the soft and collinear limit
of this particle, which is resolved in this case, have to be taken
into account for the single-unresolved contribution. Among those terms the
poles do not cancel, because the subtraction terms in ${\hat
  \sigma}^{\mathrm{RR}}_{ab}$ coincide with the ones in ${\hat
  \sigma}^{\mathrm{RV}}_{ab}$ and ${\hat  \sigma}^{\mathrm{F}_2}_{ab}$
only in the limits, but off the limits they do not. 
This is due to the fact that the subtraction terms in ${\hat
  \sigma}^{\mathrm{RR}}_{ab}$ are minimal in the sector variables, but not in the physical variables.
By comparing terms of similar type, corrections to the double-real
subtraction can be identified.
\begin{floatingfigure}[r]
 \includegraphics[scale=.75]{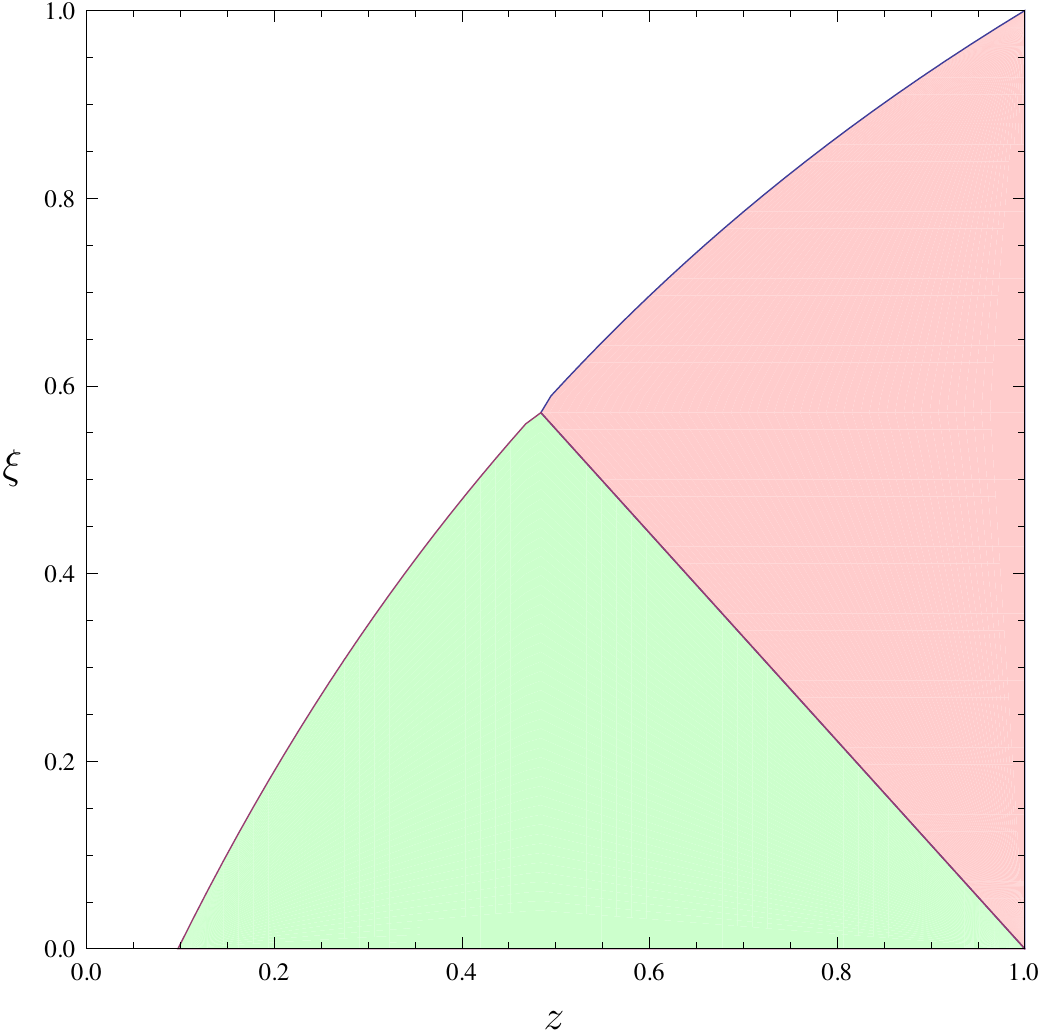}
\caption{\label{fig:Region}The subtraction term in ${\hat
    \sigma}^{\mathrm{RR}}_{ab}$ is integrated over the green region of
  phase space. It should cancel the same subtraction term in ${\hat
  \sigma}^{\mathrm{F}_2}_{ab}$, which is integrated over the red and
  green region. Therefore, the missing part (red) has to be added to  ${\hat
    \sigma}^{\mathrm{RR}}_{ab}$. Definitions of the variables
  $\{z,\xi\}$ are given in the text.}
\end{floatingfigure}
For example, the pole term due to a collinearity in the intial state
and a soft subtraction for the resolved particle should cancel between  
${\hat  \sigma}^{\mathrm{RR}}_{ab}$ and ${\hat
  \sigma}^{\mathrm{F}_2}_{ab}$. Mapping the soft variable of the
unresolved parton in the double-real contribution to $z$, the convolution variable in the collinear counter\-term, and denoting the soft variable
of the resolved particle in both contributions by $\xi$, a direct comparison is
possible. The subtraction term is the same in both cases but the
integration region is different. As depicted in figure
\ref{fig:Region} the term in ${\hat  \sigma}^{\mathrm{RR}}_{ab}$ is just
integrated in the green region, whereas it should also be integrated in
the red region to match the contribution in the collinear
counterterm. The correction can be added to the subtraction term.\\
In a similar way all corrections are identified. 
\section{Summary and Outlook}
In this proceedings, we described the general subtraction scheme STRIPPER and
explained how it can be generalized to arbitrary multiplicities by
reformulating it in the 't Hooft-Veltman regularization scheme. The
main idea was to isolate single-unresolved and double-unresolved
contributions and get them finite separately. This was achieved by
adding corrections to the integrated subtraction terms in the
double-real contribution to the NNLO cross
section. Then momenta of resolved particles were put
to four dimensions. By replacing spin correlated splitting functions by
their azimuthal averaged counterparts, we got rid of spin correlations
between four-dimensional matrix elements and $d=4-2\ep$ dimensional vectors. All
matrix elements are just needed in four dimensions.
Summing up, the extended formulation of STRIPPER furnishes a general
framework to  calculate  cross sections at NNLO in perturbative QCD.

\newpage
\section*{Acknowledgments}

\noindent

This research was supported by the German Research Foundation (DFG)
via the Sonderforschungsbereich/Transregio SFB/TR-9 ``Computational
Particle Physics''. 


\end{document}